\documentclass[twocolumn,
superscriptaddress,
 amsmath,amssymb,eqnarray
 aps,
]{revtex4-1}

\usepackage{graphicx}
\usepackage{dcolumn}
\usepackage{bm}

\begin{document}

\title{An information entropy interpretation of photon absorption by dielectric media}

\author{Sung Wook Han}

\author{Kim Myung-Whun}
\email{corresponding author: mwkim@jbnu.ac.kr}
\affiliation{Department of Physics, Chonbuk National University, 54896 Jeonju, South Korea}%
\affiliation{Institute of Photonics and Information Technology, Chonbuk National University, 54896 Jeonju, South Korea}

\date{\today}

\begin{abstract}
We measured photon absorption in dielectric media and proposed the photon-version Beer--Lambert\textquoteright s law to quantify the absorption. We used a Hong--Ou--Mandel interferometer and 810 nm twin-photons. We found that the depth ratio of the null point in the interference patterns of the interferometer agreed with the classical transmittance of the samples. We established a statistical model of the photon absorption process and proposed an information entropy interpretation to understand the meaning of the Beer--Lambert law. Comparisons of the results of the photon absorption experiments with classical experiments demonstrate the validity of our model and interpretation.
\end{abstract}


\maketitle


Research utilizing photons has recently bloomed in communication\cite{Duan,Chiribella,Greganti,Yuan} and in spectroscopy\cite{Rezus,Davis}. Photons can be transmitted over some distance through a dielectric medium such as a photonic circuit, optical fiber, or biological tissue. Although photon loss caused by the medium is negligible in usual circumstances, some amount of loss is inevitable when photons are transmitted repeatedly or over long distances \cite{Brassard,Hong1}. Such loss is primarily due to absorption by the dielectric material, with increasing photon loss accompanying better-absorbing dielectric materials. Related formulas have long been established, with Beer--Lambert\textquoteright s law one of the most well-known for quantifying absorption \cite{Mayerhöfer}; this law is based on the assumption that the intensity of an electromagnetic wave decreases at a constant rate as the wave travels a certain distance through a given medium. Thanks to its simplicity, it is widely used in materials science, chemistry, and biology as the basis of modern absorption spectroscopy.

A curiosity can be raised by considering whether the simple and useful Beer--Lambert\textquoteright s law can also be used to quantify the loss of photons. Even if the law is valid for this purpose though, the more important question is what kind of information about the photons we can get from the Beer--Lambert\textquoteright s law. The reason we need to ask this question is that while the physical meaning of absorption is well understood from a material point of view with the help of classical physics and quantum mechanics, on the other hand comprehensive approaches that seek the meaning of light absorption from the light\textquoteright s (i.e. photon\textquoteright s) point of view are seemingly rare. This question is important for those studies which concern the loss of the information more than energy or power. It is true that a number of theoretical studies based on modern quantum field theory have provided sophisticated mathematical descriptions of photon absorption in dielectric media \cite{Huttner,Barnett}. However, even in these approaches, absorption coefficients are merely treated as a material parameter describing  the attenuation of the light field strength \cite{Jeffers,Zhou}, and thus they do not clearly provide the implications of absorption from the perspective of the photon.

In this paper, we present a statistical description to understand the meaning of absorption process of photons by the dielectric medium in terms of information theory. Experimentally, we generated twin photons using spontaneous parametric down-conversion (SPDC) and passed them through dielectric media. We then compared the statistical characteristics of the twin photons by counting the number of photon pairs using a Hong--Ou--Mandel (HOM) interferometer. From the results, we are able to explain photon loss by the absorptive material based on information entropy and find that absorption coefficients can be understood as changes in the entropy.
 
Before detailing our experiment, we look at a classic Beer--Lambert style absorption experiment. In classical setups, the intensity of the light source is typically measured first as the reference, and then a sample is placed in the middle of the light path to determine the extent of light intensity reduction by the sample. Transmittance is obtained by calculating the intensity ratio of the reference light and the sample light. One assumption in this process is that the light source does not change during the two consecutive measurements. In the proposed experiment, that utilizes twin photons as the light source, such an assumption is not necessary: we can use one photon of the pair as the reference and the other photon as the sample probe, which allows us to compare the statistical characteristics of the two with high accuracy.

\begin{figure}
    \centering
    \includegraphics[width=8.2cm,bb=50 90 530 760]{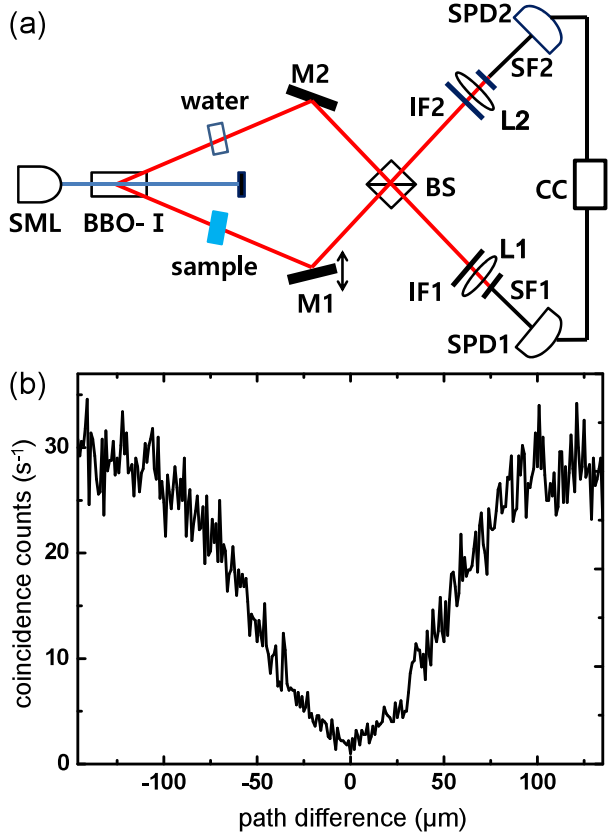}
    \caption{(a) Experimental SPDC-HOM setup for generating two photon pairs, with the following elements. SML: 100 mW, 405 nm CW single-mode laser; BBO-I: type-I β-barium borate crystal; M1/M2: mirrors (M1 with position control unit); BS: beam splitter with a transmittance/reflectance ratio of 50/50 for 810 nm; IF1/IF2: interference filters for 810 nm $\pm$ 10 nm; L1/L2: focusing lenses; SF1/SF2: single-mode fibers ; SPD1/SPD2: single-photon detectors; CC: coincidence counter. The water and samples were contained in 1-cm-thick quartz cuvettes. (b) a free-space measurement of the Hong--Ou--Mandel (HOM) dip.}
    \label{fig:1}
\end{figure} 

Figure 1(a) depicts the present experimental setup used for our absorption measurements of twin photons. The twin photons were generated by a type-1 SPDC process and counted with an HOM interferometer. For SPDC, we used a 405-nm continuous wave (CW) laser with a type-1 $\beta$-barium borate (BBO) crystal. The transmittance/reflectance ratio of the beam splitter was 50:50, and 810-nm pass filters were placed in front of the single-photon counters. Figure 1(b) shows the HOM interference curve of twin photons in the absence of any dielectric medium except air; we call this the free-space curve. The interference null, also called as the HOM dip resulted in $95 \%$ visibility. This indicates that the two-photon correlation function is smaller than one (i.e. $g^{(2)} < 1$) at zero path difference ($\delta =0$) and means that the number of photon pairs can be counted with high accuracy. To the setup, we inserted reference and sample materials in the paths of each beam before the beam splitter. In the reference path, we inserted a quartz cuvette (1 cm $\times$ 1 cm $\times$ 5 cm) filled with deionized water, and in the sample path, we inserted an identical cuvette filled with aqueous solutions of copper sulfate pentahydrate. Five solutions of differing molal concentrations were prepared: 0.005 M, 0.01 M, 0.02 M, 0.03 M, 0.06 M and 0.09 M. 

Figure 2(a) shows the coincidence counts of the samples. The curve labeled \textquotedblleft water\textquotedblright  represents the measurement with two deionized water cuvettes in both paths; we will call this combination the \textit{water--water} case. The other curves represent the \textit{water--sample} cases, where one cuvette contained deionized water and the other contained copper sulfate solution. The plots in Fig. 2(a) indicate that as sample concentration increased, coincidence count decreased. The solid lines overlaid on the coincidence curves are the $g^{(2)}$ functions of the ideal situation. The fitting lines for the water–sample cases were obtained by multiplying a constant to the $g^{(2)}$ function of the water--water case.

\begin{figure}
    \centering
    \includegraphics[width=8.2cm,bb=20 0 540 800]{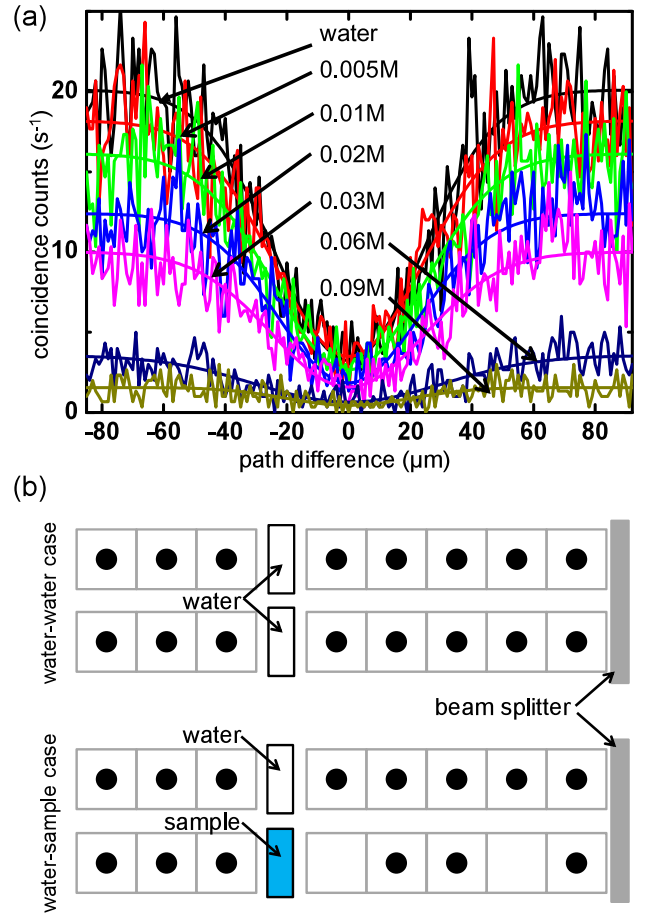}
    \caption{(a) HOM dips of the water--water and water--sample (copper sulfate pentahydrate solution) cases with fitting curves (solid lines). (b) Model sketch of the photon-pair stream in the HOM setup shown in Fig. 1(a). The upper and lower halves represent the water--water and water--sample cases, respectively.}
    \label{fig:2}
\end{figure}    

We then conducted a thought experiment. The upper sketch in Fig. 2(b) is a snapshot of the photon streams before the beam-splitter in the water--water case. The two streams are directed towards the beam-splitter, and the path-length difference between the streams is zero. In this figure, the horizontal direction is the photon propagation direction, each square box represents the time resolution of the detector, and the solid circles each represent one 810 nm photon. Photons of other wavelengths are ignored because they will be blocked by the 810 nm pass filters. Note that the circle does not represent a photon as a real quantum mechanical object (the photon wave-function). It should rather be considered as an electrical signal pulse to be generated by a photon at each detector within the time resolution. The photon pairs in the upper panel are maintained before and after the cuvettes since the photons experience no absorption in the water--water case. The lower sketch in Fig. 2(b) shows the water--sample case, where the photon stream passing through the water cuvette is unchanged but the one passing through the sample cuvette experiences photon loss, which reduces the number of photon pairs. This model suggests that the number of photon pairs contains information about absorption. We assume that photon loss due to reflection is the same in both cases and we ignore the reflection loss in the sketch.

If we neglect the accidental coincidence, the coincidence count is zero when the path difference is zero (i.e. $\delta = 0$). When $\delta$ is large, on the other hand, every photon pair arriving at the beam splitter simultaneously should yield a coincidence count. The difference of the coincidence counts between $\delta = 0$ and $\delta >> 0$ is the same as the depth of the coincidence null, and it should be the number of photon pairs arriving at the detectors for the measurement time. According to the model in Fig. 2(b), we can estimate the probability of photon transmission by counting the number of boxes not containing photons. Such counting can be performed by single-photon detectors and the coincidence counter. Coincidence detection by the two photon-counters can be considered analogous to opening the two boxes at the same time and checking the number of photons therein at every time resolution.

We denote the average number of photon pairs obtained per time resolution as $\langle n \rangle$, and we define the photon transmittance based on $\langle n \rangle$. The number of photon pairs in the actual experiment is measured for a relatively long time (e.g., one second), which should be an integer number ($N$) times $\langle n \rangle$; this can be thought of as classical light intensity. On the other hand, the $N\langle n \rangle$ value also corresponds to the difference between the coincidence counts at $\delta >> 0$ and the coincidence counts at $\delta = 0$ in the HOM interferometer. We call this difference the HOM dip depth ($D$). Let $D_w$ be the dip depth value in the water--water case and $D_s$ in the water--sample case. If the signal measurement times of both cases are the same, then the ratio of the light intensities passing through the sample and through the water can be written as
\begin{equation}
\frac{I_s}{I_w} = \frac{N \langle n \rangle_s}{N \langle n \rangle_w} = \frac{D_s}{D_w}.
\end{equation}
From Eq. (2), we can derive a photon version of the classical Beer--Lambert\textquoteright s law:
\begin{equation}
\alpha \equiv -\ln \frac{D_s}{D_w}.
\end{equation}
Photon absorption coefficient ($\alpha$) results obtained from Eq. (3) are shown in Fig. 3 (circles). For comparison, we measured the transmittance of the water and samples with a conventional spectrometer, and using the classical Beer--Lambert\textquoteright s law, we obtained the classical absorbance and plotted the results in Fig. 3 (squares). The two results are consistent, demonstrating that our model for photon absorption and the formula in Eq. (3) are valid.

\begin{figure}
    \centering
    \includegraphics[width=8.2cm,bb=40 30 670 550]{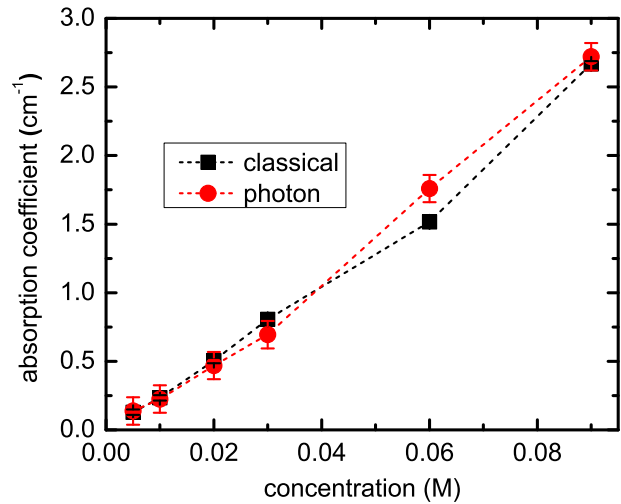}
    \caption{(squares) Classical absorption coefficient at 810 nm: Transmittance was measured with a classical spectrometer, and the absorption coefficient was calculated from the classical Beer--Lambert\textquoteright s law. (circles) Photon absorption coefficient: Transmittance was measured with the HOM interferometer in Fig. 1(a), and the absorption coefficient was calculated from the proposed photon version of the Beer--Lambert\textquoteright s law in Eq. (3)}
    \label{fig:my_label}
\end{figure} 

Based on the photon-counting model, we suggest the meaning of photon absorption from a statistical point of view. If we assume that the absorption of photons by a sample is a stochastic event, then we can understand the event as a change in the information entropy of the photon stream. In our setup, the photon detector can recognize the incident photon as a 1 or 0 (presence or absence) within the time resolution; 1 means one photon and 0 means no photons. By imagining a photon stream passing through the water and sample cuvettes for a certain time ($N\times$ time resolution), we can consider this process as sequential N events of opening two pairs of boxes at the same time. Note that the act of opening two pairs of boxes is an independent event and causes no effect on the opening of the next two boxes. We define $S = \left\{s_i\right\}$ and $W = \left\{w_i\right\}$ as the set of events when the photons pass through the sample cuvette and the water cuvette, respectively. If we open a pair of boxes among the $N$ pairs of boxes, we can observe only four possible joint events $(s_i, w_i)$: $(0, 0)$, $(0, 1)$, $(1, 0)$, or $(1, 1)$. Every event is independent.

From the stochastic point of view, we now define photon transmittance as the ratio of the (1, 1) event probability in the water--sample case to the same probability in the water--water case. Since the former is more complicated, we first calculate the transmittance of the water--sample case and then apply the same idea to the water--water case. According to the sketch in Fig. 2(b), we can estimate the probability that both the box following the water and the box following the sample contain photons simultaneously. We can quantify the amount of information needed to explain the event through a joint information entropy $H(S, W)$.

The $H(S,W)$ can be calculated by using the joint probability $p(s_j,w_i)$, which is the joint probability of a particular $s_j$ event in set S occurring simultaneously with a particular $w_i$ event in set W. With this joint probability, $H(S,W)$ can be written as 
\begin{equation}
\begin{split}
H(S,W)&=-K\sum_{s_j\in S}\sum_{w_i \in W}p(s_j,w_i)\log_2 ⁡p(s_j,w_i)\\
       &=H(1,1)+H(0,1)+H(1,0)+H(0,0)\\
       &=H(1,1)+H(e)
\end{split}
\label{eq:Eb2}
\end{equation}
where $K$ is a positive constant. We employ binary logarithms because the information we are handling is bit information indicating the presence or absence of a photon. To change the information unit from bit to nat, we convert the binary logarithms to natural logarithms by multiplying $\ln2$ to the entropy (i.e. $k = K\ln2$). In Eq. (4), $H(S = 1, W = 1)$ is the entropy of the event in which the photon that passes through the water is paired with the photon that passes through the sample, and $H(e)$ is the entropy of all other events. From the viewpoint of coincidence count, $H(S = 1, W = 1)$ corresponds to the information of the photon-pair coincidence event $(x_1)$, and $H(e)$ is the information of all other events $(x_2)$. We define the number of photon pairs counted in the $x_1$ event as $n(x_1)$, and that in the $x_2$ event as $n(x_2)$. Note that $n(x_1) = 1$ and $n(x_2) = 0$. If $p_1$ and $p_2$ are the probability of each event, then we can write $H(S = 1, W = 1) = H(p_1)$ and $H(e) = H(p_2)$. The joint entropy $H(S, W)$ of two independent events is equal to the sum of the entropy of the $x_1$ event and the entropy of the $x_2$ events. Thanks to this relation, we can consider the measurement of the photon pair as a Bernoulli process, i.e. $H(p_1 )+H(p_2 )=-k\sum_{i=1,2} p_i \ln p_i$.

We have no accurate a priori information about the probability. As the only detectable information from the experiment is the number of photon pairs, we need to find a way to figure out the probability from the photon-pair count. According to information theory, even when probability $p_i$ cannot be known directly, it can be estimated from accessible experimental information along with the fact that the sum of the probabilities of event occurrence should be 1 ($\sum_i p_i =1$) \cite{Jaynes}. In our experiment, we can obtain the average number of photon pairs as $\langle n \rangle$ ($=\sum_{i=1,2} p_i n(x_i)$) at each resolution. Based on these pieces of information, we can write the following auxiliary function:
\begin{equation}
\begin{split}
  F=&-k\sum_{i=1}^2 p_i \ln p_i +\lambda\left(\sum_{i=1}^2 p_i -1\right) \\ 
    &+\mu\left(\sum_{i=1}^2 p_i n(x_i)-\langle n \rangle\right)
\end{split}
\end{equation}
Here, $\gamma$ and $\mu$ are Lagrangian multipliers. According to the principle of maximum entropy inference, the best probability estimate is to maximize the $F$ function, i.e., $\partial F/\partial p_i=0$, with $-k\ln p_i -k+\gamma+\mu n(x_i )=0$ satisfying this condition. We then set $-(\mu/k)\equiv \beta$, $(1-\gamma/k)\equiv\lambda$, giving $p_i = \exp(-\lambda - \beta n(x_i ))$. Using this probability, the average photon number $\langle n \rangle$ can be calculated as
\begin{equation}
\langle n \rangle = \sum_{i=1}^2 p_i n(x_i )=-e^{-\lambda}\frac{\partial}{\partial \beta} \sum_{i=1}^2 e^{-\beta n(x_i)} =-\frac{\partial}{\partial \beta} \ln Z
\end{equation}
where we use $\lambda = \ln Z$ and $Z=\sum_{i=1,2} e^{-\beta n(x_i)}$. Since we assume that $n(x_1) = 1$ and $n(x_2) = 0$, the partition function satisfies the relation $Z=1+\exp^{-\beta}$, in which $\beta$ can be regarded as the inverse temperature in thermodynamics point of view.

We now express the photon transmittance defined in Eq. (2) in terms of the probability. Since $\langle n \rangle = p_1 = \exp(-\beta-\lambda)$, Eq. (2) can be re-written as
\begin{equation}
\frac{\langle n \rangle_s}{\langle n \rangle_w} = \frac{p_{1s}}{p_{1w}} =\frac{e^{-\beta_s-\lambda_s}}{e^{-\beta_w-\lambda_w}} =\frac{Z_w}{Z_s}e^{-(\beta_s-\beta_w)}=e^{-\alpha},
\end{equation}
where $\alpha=(\beta_s-\beta_w)+(\lambda_s-\lambda_w)$. From the relationship between entropy and the partition function, we know that $H / k = ln Z + \beta \langle n \rangle$. If we consider the limiting case where absorption is weak, i.e. $\langle n \rangle \simeq 1$, then we can assume that $\beta \langle n \rangle \simeq \beta$. The assumption leads to the relation $e^{-\alpha} \simeq e^{-(H_s-H_w)/k}$. We can write this into a simpler form as
\begin{equation}
\alpha \simeq \frac{\Delta H}{k},
\end{equation}
which means that absorption coefficient $\alpha$ approximately equals the increase of the information entropy due to the loss in the photon stream.
  
In conclusion, we presented a method for quantifying the photon absorption of a dielectric medium by analyzing Hong--Ou--Mandel interference patterns. We demonstrated that the photon absorption obtained by our method is in good agreement with classical absorption from the conventional method, which shows that our photon version of the Beer--Lambert\textquoteright s law based on the presented statistical model is valid for quantifying photon absorption. Our model shows that photon absorption is related to the change in information entropy of photon pairs passing through the sample. We measured absorption using less than 20 photons per second, indicating that the photon version of the Beer-Lambert law can be applied in experiments where extremely dim light is used.

We acknowledge the support of the National Research Foundation of Korea (grant number: 2016R1D1A1B03934648).

\end{document}